\documentclass[aps,prl,twocolumn,superscriptaddress,floats]{revtex4}
\usepackage{graphicx}
 \usepackage{bm}

\usepackage{color}
\usepackage[percent]{overpic} 
\definecolor{blue}{rgb}{0,0.0,0.9}
\definecolor{green}{rgb}{0.0,0.9,0.0}
\begin{document}

\title{How the velvet worm squirts slime}
\author{Andr\'{e}s Concha}
\affiliation{School of Engineering and Sciences, Adolfo Iba\~{nez} University, Diagonal las Torres 2640, Pe\~{n}alolen, Santiago, Chile}

\author{Paula Mellado}
\affiliation{School of Engineering and Sciences, Adolfo Iba\~{nez} University, Diagonal las Torres 2640, Pe\~{n}alolen, Santiago, Chile}

\author{Bernal Morera-Brenes}
\affiliation{Laboratorio de Gen\'{e}tica Evolutiva, Escuela de Ciencias Biol\'{o}gicas, Universidad Nacional, 86-3000  Heredia, Costa Rica}
\author{Cristiano Sampaio}

\affiliation{Universidade de Sao Paulo, Sao Paulo, SP, Brazil}

\author{L. Mahadevan}
\affiliation{School of Engineering and Applied Sciences, Harvard University, Cambridge, MA 02138,USA}
\affiliation{Department of Physics, Harvard University, Cambridge, MA 02138}

\author{Juli\'{a}n Monge-N\'{a}jera}
\affiliation{Tropical Biology, Universidad de Costa Rica, 2060 San Jos\'{e}, Costa Rica}


\maketitle

\textbf{ The rapid squirt of a proteinaceous slime jet endows the ancient velvet worms (Onychophora) with a unique mechanism for defense from predators and for  capturing prey by entangling them in a disordered web that immobilizes their target. However, to date neither qualitative nor quantitative descriptions have been provided for this unique adaptation. Here we investigate the fast oscillatory motion of the oral papillae and the exiting liquid jet that oscillates with frequencies $f\sim 30-60$ Hz. Using anatomical images, high speed videography,  theoretical analysis and a physical simulacrum we show that this fast oscillatory motion is the result of an elastohydrodynamic instability driven by the interplay between the elasticity of oral papillae and the fast unsteady flow during squirting.  Our results demonstrate how passive strategies can be cleverly harnessed by organisms, while suggesting future oscillating micro-fluidic devices as well as novel ways for micro and nano fiber production using bioinspired strategies.}

Rapid motions in nature are seen in a variety of situations associated with escape and predation. Extreme examples include the  chameleon tongue that uses an unusual muscle-spring configuration to capture prey \cite{muller2004power} and the venus flytrap that stores elastic energy and uses an instability for the rapid closure of its leaf \cite{forterre2005venus}. The velvet worm is an unusual example of how an organism projects itself by squirting a jet of slime in an oscillatory fashion, not only for capturing prey, but also for defense \cite{belt1888naturalist,bouvier1905monographie,joseph1928,alexander1957notes,alexander1958fierce}.  Despite having been a  subject of study for over a century, the mechanism underlying the rapid oscillatory squirting of slime by the velvet worm remains a mystery \cite{morera2010new,haritos2010harnessing,belt1888naturalist}. Indeed, Darwin even hypothesized the creation of the disordered web as a potential origin for the evolution of spider webs \cite{moseley1874structure},

To capture the dynamics of the squirting process, we filmed several worm attacks (see Supplementary Information for movies). In Fig.\ref{FIG1}a-d, we show a series of snapshots of an attack recorded using high speed imaging (480 fps), with the  average duration of a squirt for all specimens being $\Delta t_{ave} =0.064\pm 0.005$ s (See Supplemental Information S2). By tracking the motion of the tip of the jet shown in Fig.\ref{FIG1} and Fig.S2, as a function of time,  we found that the typical jet speed $v\sim 3-5$ m$/$s. Furthermore, we see that the squirt does not remain oriented but instead sprays an entire region as shown in Fig.\ref{FIG1}e. These measurements raise the natural question of the spatio-temporal evolution of the liquid jet and its control by the worm. 

Actively controlled muscular action has long been invoked as the $natural$ explanation for the spectacular way in which these worms quickly weave their web, and continues to be the favored mechanism  \cite{morera2010new}. However, papillar oscillations are fast (Fig. \ref{FIG1} and Supplementary Information) in comparison with any other motion of the worm ($f_{papilla}/f_{walking}\sim 30-60$), and with known time 
scales ($\sim 0.5$ sec) for the fastest muscles in the worm \cite{hoyle1979neuromuscular}, suggesting a conceptual difficulty with this hypothesis.  Therefore, we examined the anatomy of the whole squirt system and surrounding tissues (Fig. \ref{FIG2} and Fig. S1). During squirting, the oral papilla extends from its folded shape to its full length  (Fig.\ref{FIG2}) of up to $L\sim 6$ mm  (Supplementary information Fig.S2). In Fig.\ref{FIG2}a we see a large reservoir region (\textbf{re}) where slime is stored, and a narrow duct that ends at the oral papilla, a syringe-like geometry that facilitates the acceleration of the slime for the fast squirt. In Fig.\ref{FIG2}a-b we show that muscle fibers in the oral papilla are similar to those found in the legs, but fewer in number. Fig.\ref{FIG2} a-b and  Fig. S1 also show that muscular fibers found in papilla tissues are consistent with their directional function, with some being annular, typical of sphincter-like systems. The relaxed papilla has an accordion shape (Fig.\ref{FIG2}c,d) that is unfolded just before the squirting process, and can thus be easily packed while also having an inhomogeneous bending rigidity, with soft spots that make papillae more pliable and susceptible to bending as slime is squirted. 

Our anatomical findings are consistent with earlier evidence that slime papilla are modified limbs \cite{budd2002palaeontological,scholtz2006evolution} with a nervous system similar to that in their legs \cite{mayer2009neural}. Detailed descriptions  
of the Onychophora muscular system \cite{hoyle1979neuromuscular} show that the fastest muscles are located in the jaw with typical twitch time scales $\sim 0.5$s,  which while fast for this primitive worm, are nearly 25 times slower relative to the papillary oscillation time scale $\sim 20$ms. Given that the legs and the papillary muscles are even slower, consistent with the primitive nature of these worms \cite{bouvier1905monographie,liu2011armoured,fortey1993case}, we are left with an obvious question - how are rapid changes in direction, that occur over a time scale of a few milliseconds  possible without the existence of any specialized rapid muscular actuation or neural control ? 

A way around this conundrum is to realize that a physical mechanism can drive the rapid and nearly chaotic oscillations of the papilla - just as a garden hose pipe develops a life of its own when water squirts out of it rapidly. Indeed, the inertial effects associated with the exiting fluid jet  drive the elastic hose pipe to flutter, a subject that has been well studied experimentally and theoretically at the macro scale \cite{paidoussis1998fluid}. In the current microscopic setting, the interplay between fluid forces and the papilla elasticity produce the characteristic oscillatory waving motion used to capture prey, and obviate the need for any fast moving controlled muscles. However, this mechanism requires fluid inertia to play a critical destabilizing role, i.e.  the ratio of inertial to viscous forces characterized by the Reynolds number $Re = vR/\nu > 1$ (where $v$ is the characteristic fluid velocity, $\nu$ the fluid kinematic viscosity, and $R$ the tube radius). In microfluidic geometries where typical sizes are small, unless the velocities are sufficiently large, inertial effects are unimportant.

Our microscopy studies (see Fig.\ref{FIG2}a) show that the squirting system has a reservoir that contracts $slowly$ driving the slime  through a small duct that runs close to the center of the oral papilla (Supplemental information S3). This geometric amplifier can lead to an increase in the speed during squirting. Our observations are in contrast with previously reported studies \cite{manton1937}, that have persisted into the modern literature \cite{barnes2009invertebrates}, where no mention of the cross section reduction has been made. Our micrographs show muscular structures (Fig.\ref{FIG2} and Fig. S1) around the slime reservoirs which are functionally consistent with the contraction of this organ. These structures resemble the design of radial tires where a fiber network is used to reinforce the wall \cite{clark1981mechanics},  consistent with detailed study  of muscular fibers at reservoir level \cite{baer2012comparative}. Measurements of the squirted volume (Supplemental information S3) and reservoir geometry show that the  contraction ratio $\delta R_{re}/R_{re}<0.03$ (with $R_{re}\sim 2$ mm) changes in about $0.1$ seconds, enough to produce speeds of $v\sim 5 $ m$/$s, so that the Reynolds number $Re \sim 2700$ (see Supplementary Information S7).  Since we do not see perfect synchronization between liquid jets coming from different papilla (Fig.\ref{FIG1}a-d and Supplementary Information S9,10),   whole body contraction as the main driving force in squirting \cite{baer2012comparative} is unlikely. 

This leads to the conclusion that the instability arises due to a competition between fluid inertia and elastic resistance. When a liquid moves steadily through a flexible pipe at small $v$, flow-induced damping prevents any oscillations from growing. For large enough $v$, centrifugal  and Coriolis forces make the pipe unstable for fluid speeds $v >v_{c}$.  In the limit when the effects of gravity  can be neglected (see Supplementary Information S6),  a simple scaling argument allows us to estimate the frequency of oscillations $f$ by balancing  the stabilizing elastic bending resistance with the destabilizing inertial forces, i.e. $E I/ \lambda^4\sim M v f/\lambda \sim Mv^2/\lambda^2$, where $E I$ is the the bending stiffness, $\lambda\sim 2 L$ is the approximate oscillation wavelength for the cantilevered papilla, $M$ is the mass density per unit length of the fluid in the pipe,  This yields  $f\sim\left(EI/M\right)^{1/2}/(2L)^2 \ $. Similarly, a typical speed scale can be estimated as $u_{0} \sim f \lambda = \left(\frac{E I}{M}\right)^{1/2}\frac{1}{L}$. For the specimen shown in Fig.S2 the oscillation frequency $f \sim 58$ Hz,  papilla length $L=6.0$ mm, papilla outer diameter $D=1.0$ mm, papilla inner diameter $d=0.5$ mm, from where $\beta\sim 0.25$. The measured frequency allows to estimate the effective Young's modulus of the papilla to be $\sim 20$ kPa consistent with measurements using two small magnets to pull the papilla that yields $E\sim 40$ kPa (Supplementary Information S5), and thence the typical speed $u_{0} \sim 0.5$ m$/$s.  The precise critical speed $v_c$ depends on boundary conditions, which for the  cantilever case gives $v_{c}\sim 2\pi u_{0}=3$ m$/$s. At this critical speed, stability is lost via a Hopf type bifurcation \cite{paidoussis1998fluid} (see SI for further details).

For unsteady flows, such as when the jet is  being accelerated inside the flexible papilla, the fluid acceleration  $dv/dt$ can destabilizes the system at even lower jet speeds (See SI numerical simulations and experiments with an unsteady synthetic system).   Our measurements show that $v\sim 3.2-5.0$ m$/$s. Therefore, even without muscular action the papilla will become unstable due to a simple physical instability \cite{paidoussis1998fluid}. 
 
To show that it is indeed possible to drive these oscillatory instabilities on small scales, we made a synthetic papilla out of a soft elastomer in the form of a flexible micro-pipe with a rectangular cross section that defines an oscillation axis for its softest bending mode. Our pipe was moulded out of PDMS with a Young's modulus $E=288$ kPa  (Supplemental information Fig. S5)  with thickness $h=1.42$ mm, width $w=1.60$ mm, inner diameter $d=0.81$ mm, and length $L=9.5$ mm. Our model system is simpler than the natural one in at least two aspects: There is no roughness along the inner part of the duct, or an external accordion like geometry.    From measurements of external diameters (Fig.\ref{FIG2}d) we found that $B=EI$ can locally change up to $1/10$ of the stiffness corresponding to a uniform soft papilla of diameter $D_{0}$, showing that the accordion like microstructure will lower $v_{c}$. Our experimental results (Fig.\ref{FIG3}, Supplementary Information and Movie $4$) show that synthetic papilla becomes unstable and oscillates when the liquid reaches a speed of $v_{c}=8.6$ m$/$s. This occurs in the same range of dimensionless parameters that for the natural organ in agreement with theoretical predictions \cite{paidoussis1998fluid} and shows that fast muscular action at the papillary level is unnecessary for oscillations.  Naturally, this experiment also suggests a prototype of a fluid driven micro-mechanical actuator \cite{whitesides2006origins,leslie2009frequency,ho1998micro}.
 
Our observations, minimal theory and a physical mimic show how velvet worms can spray a web rapidly to entangle their prey without resorting to a complex neuromuscular control strategy. Instead, they can simply harness the   instability associated with rapid flow through a long soft nozzle that causes it and the exiting jet to oscillate. As this jet solidifies rapidly, it entraps the object dynamically. While there is a superficial similarity to spider webs, we see that here dynamics is of essence. Our findings and synthetic model might also pave the way to self supported flexible micro-fluidic devices that can take advantage of similar instabilities in order to  produce a variety of products such as micro drops \cite{kaufman2012structured,kong2012droplet,PhysRevLett.112.054501}, and non woven fiber structures  \cite{mellado2011simple}.

\textbf{Acknowledgement}
A.C. and P.M. acknowledge partial support from Conicyt PAI $79112004$.  A.C and P.M. were also partially supported by Fondecyt  grants $11130075$ and $11121397$ respectively. L.M. acknowledges the support of the MacArthur Foundation. A.C would like to thank Zhiyan Wei for sharing his insights about instabilities in flexible fibers and related phenomena. We also acknowledge S. Rica, J. Poupin, T. Ledger, and B. Gonzalez for conversations and training.

\textbf{Competing Interests} The authors declare that they have no competing financial interests.
 
 \textbf{Author Contributions} AC - brought together the collaboration,  designed and built experiments,  performed imaging and numerical simulations. PM, AC performed microfluidic experiments,  carried out the fluid-dynamics analysis. LM - proposed the link between the biological problem and the artificial one, suggested the physical mechanism.   BM, JN, CS - provided the worms, performed and interpreted microscopy, carried out the biological research. All authors contributed writing the paper.

\textbf{Correspondence} Correspondence should be addressed to andres.concha@uai.cl or julian.monge@ucr.ac.cr.


\begin{figure*}
\begin{center}
\begin{overpic}[scale=.60]{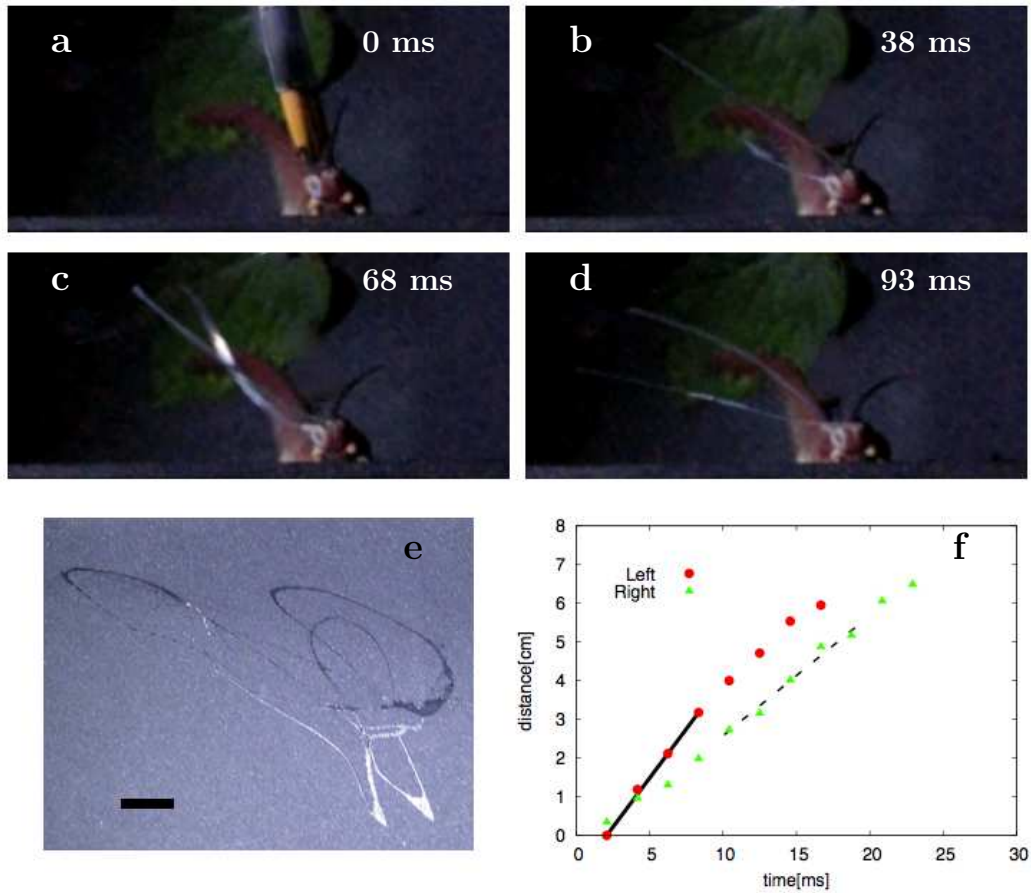}

\put(5,82){\color{white}\Large \bf{a}} 
\put(35,82){\color{white}\large \bf{0 ms}} 

\put(55,82){\color{white}\Large \bf{b}} 
\put(85,82){\color{white}\large \bf{38 ms}} 

\put(5,59){\color{white}\Large \bf{c}} 
\put(35,59){\color{white}\large \bf{68 ms}}
 
\put(55,59){\color{white}\Large \bf{d}} 
\put(85,59){\color{white}\large \bf{93 ms}} 

\put(39,33){\color{black}\Large \bf{e}} 
\put(92,33){\color{black}\Large \bf{f}} 

\end{overpic}
\end{center}
\caption{\label{FIG1} \textbf{Worm attack.}
Giant red velvet worm $\it{Peripatus}$ $\it{solorzanoi}$ used to record the squirting process. Full body length $\sim17.5$ cm. In panel \textbf{a} a soft paintbrush used to
activate its attack is shown, and it was digitally removed in the others snapshots for clarity.
$\mathbf{a-d,}$ Panels shows different stages of the attack recorded at 480 fps. The active part of the attack is completed in $\Delta t_{squirt}\sim 65$ ms.   
$\mathbf{e,}$ Slime pattern generated on a wall of the foam tunnel used to keep worms at focal distance (scale bar, 1cm). Three or more oscillations of the slime jet is the typical outcome.   
$\mathbf{f,}$  Shows liquid jet tip position as a function of time for the squirt (data taken from white specimen described in supplementary information Fig.S2). For this data we used two cameras: one at 30 fps and the other at 480 fps. The two cameras configuration allowed us to compute the jet velocity.  The solid line corresponds to a squirt speed of $v=5.0$ m$/$s, and the dotted line to a squirt speed of $v=3.2$ m$/$s. Solid dots and green squares correspond to left and right papilla respectively.}
\end{figure*}

\begin{figure*}
\begin{center}
\begin{overpic}[scale=.60]{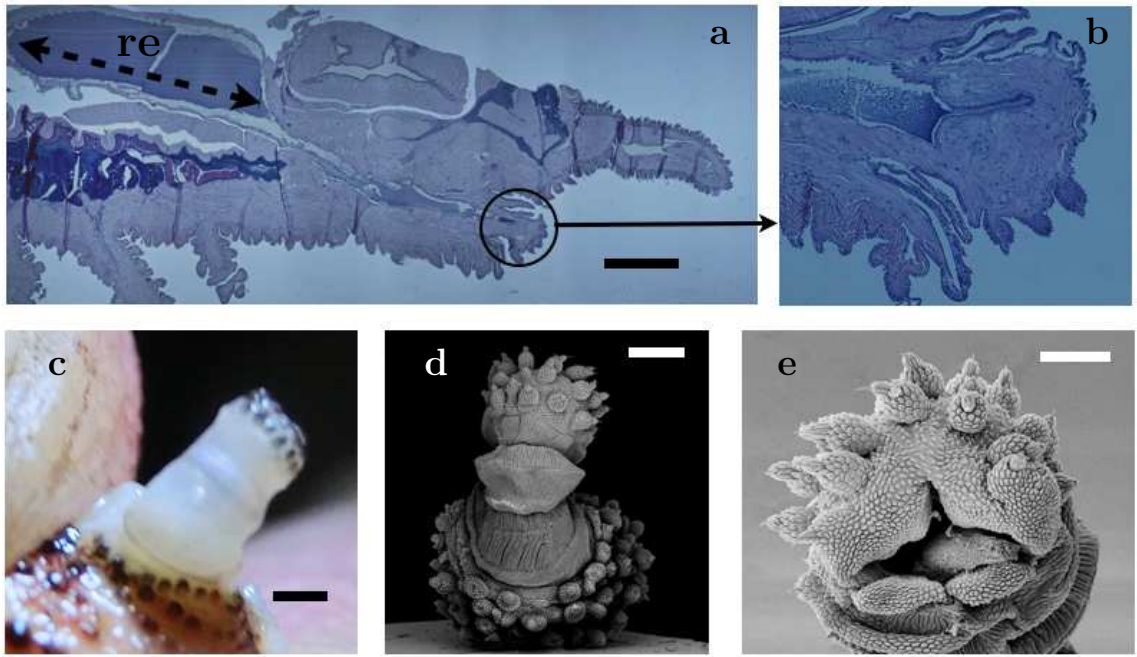}

\put(62,54){\color{black}\Large \bf{a}} 
\put(10,53){\color{black}\LARGE \bf{re}} 
\put(95,54){\color{black}\Large \bf{b}} 
\put(4,25){\color{black}\Large \bf{c}} 

\put(37,25){\color{white}\Large \bf{d}} 
\put(68,25){\color{black}\Large \bf{e}} 

\end{overpic}
\end{center}
\caption{
\label{FIG2}\textbf{Squirt system and Papilla structure:} 
$\mathbf{a,}$  Hematoxylin eosin stained cut of a $\it{Peripatus}$ $\it{solorzanoi}$.  The squirt system is composed by a large slime reservoir, \textbf{re}, which length is depicted by a dotted arrow. The reservoir has a large diameter compared with the diameter of the duct that connects it with the oral papilla  (black circle) shown in \textbf{b} (scale bar, 4mm).
$\mathbf{b,}$ Longitudinal cut of an oral papilla of a $\it{Peripatus}$ $\it{solorzanoi}$. Its wrinkled surface, and sphincter like tissues surrounding the inner duct are shown. The black substance inside the duct corresponds to remnants of the slime.
$\mathbf{c,}$ Structure of a fresh oral papilla (red $\it{Peripatus}$ $\it{solorzanoi}$) when deployed. Its semi-transparent and wavy structure is apparent (scale bar, 1mm). 
$\mathbf{d,}$ Oral papilla accordion structure of a $\it{Epiperipatus}$ $\it{acacioi}$ (scale bar, $200\mu$m). In this case the hinged accordion structure is resolved by SEM microscopy.
 $\mathbf{e,}$ Opening of the oral papilla $D\sim 350 \mu$m, and  $d\sim 200 \mu$m (scale bar, $100\mu$m). 
 } 
\end{figure*}

\begin{figure*}
\begin{center}
\begin{overpic}[scale=.5]{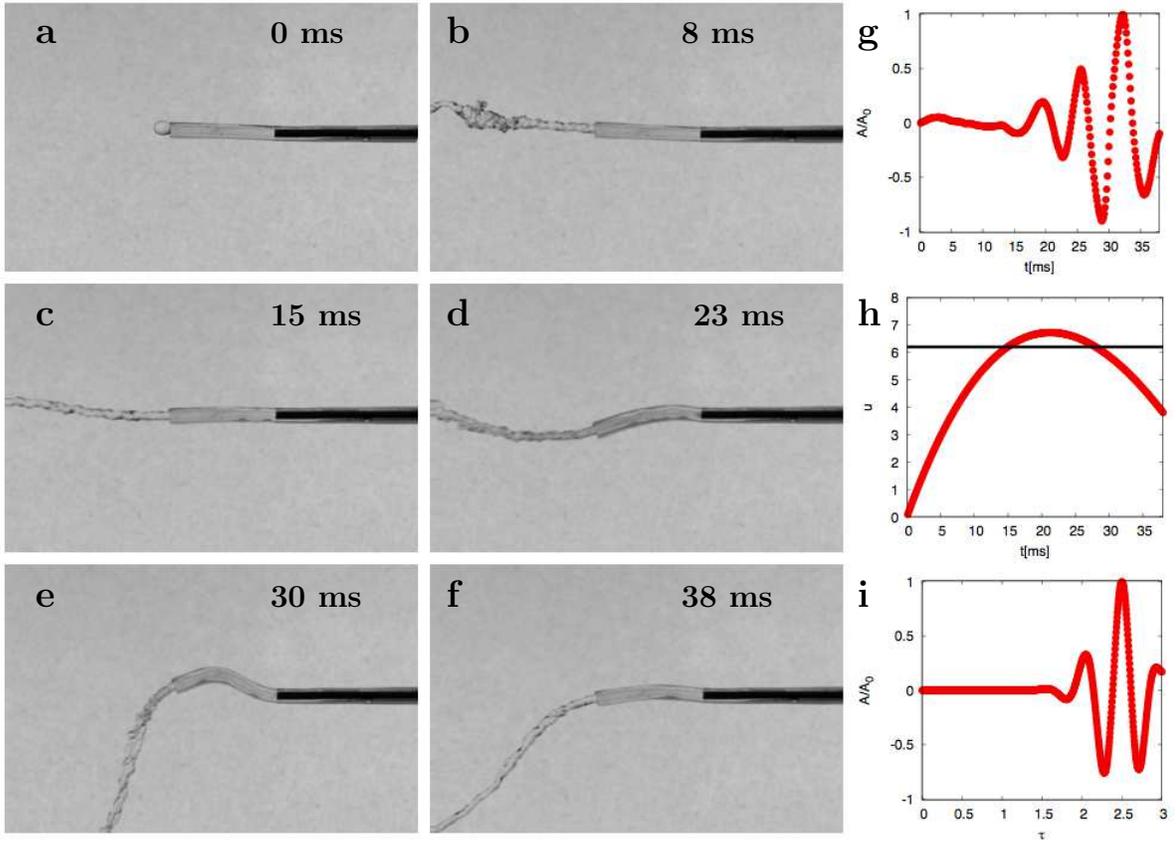}

\put(3,68){\color{black}\Large \bf{a}} 
\put(23,68){\color{black}\large \bf{0 ms}} 
\put(38,68){\color{black}\Large \bf{b}} 
\put(58,68){\color{black}\large \bf{8 ms}} 
\put(3,44){\color{black}\Large \bf{c}} 
\put(23,44){\color{black}\large \bf{15 ms}} 
\put(38,44){\color{black}\Large \bf{d}} 
\put(58,44){\color{black}\large \bf{ 23 ms}} 

\put(3,20){\color{black}\Large \bf{e}} 
\put(23,20){\color{black} \large\bf{30 ms}} 
\put(38,20){\color{black}\Large \bf{f}} 
\put(58,20){\color{black}\large \bf{38 ms}}

\put(73,68){\color{black}\Large \bf{g}} 
\put(73,44){\color{black}\Large \bf{h}} 
\put(73,20){\color{black}\Large \bf{i}} 

\end{overpic}
\end{center}
\caption{\label{FIG3} \textbf{Synthetic papilla:} 
$\mathbf{a-f}$ A soft elastic cantilevered tube made out of PDMS becomes unstable as the fluid flowing through it increases its speed(See Supplementary Information for Movie). Its height is  $h=1.42$ mm, width $w=1.60$ mm,length $L=9.5$mm and hole diameter 
 is $0.81 $ mm.  The fluid used in this experiment was water. 
 $\mathbf{g,}$ The vertical motion of a point at the center line of the hose close to the tip, as fluid speed is increased shows that oscillations develop and grow.
 $\mathbf{h,}$ During an emptying cycle, the fluid speed varies as a function of time, with the dimensionless speed $u\in$ $[0.0,6.9]$, with $u=v/u_{0}$. Horizontal black line depicts the theoretical threshold for instability in the case of steady flow (see SI for details).
  $\mathbf{i,}$ Numerical simulations of the squirting dynamics associated with the dynamics of emptying shown in panel \textbf{h} using the governing equations (see SI for details).
 }
\end{figure*}

\end{document}